\documentclass[english,superscriptaddress,showpacs,showkeys]{article}

\usepackage{graphicx}
\usepackage[utf8]{inputenc}
\usepackage[T1]{fontenc}
\usepackage{geometry}
\usepackage{amsmath}
\usepackage{array}
\usepackage[british]{babel}
\newcommand{\p}{\bar}
\usepackage{authblk}
\DeclareGraphicsExtensions{.png}
\usepackage{cite}

\begin{document}
\title{Local implementations of non-local quantum gates in linear entangled channel}

\author[1]{Debashis Saha}
\author[2]{Sanket Nandan}
\author[1]{Prasanta K. Panigrahi$^{\footnote{pprasanta@iiserkol.ac.in}}$}

\affil[1]{\it Indian Institute of Science Education and Research Kolkata, Mohanpur Campus, Nadia-741252, West Bengal, India}
\affil[2]{\it Indian Institute of Science Education and Research Pune, Dr. Homi Bhabha Road, Pashan, Pune-411008, Maharashtra, India}

\date{}
\maketitle

\begin{abstract}

In this paper, we demonstrate $n$-party controlled unitary gate implementations locally on arbitrary remote state through 
linear entangled channel where control parties share entanglement with the adjacent control parties and only one of them shares entanglement with 
the target party. In such a network, we describe the protocol of simultaneous implementation of controlled-Hermitian 
gate starting from three party scenario. We also explicate the implementation of three party controlled-Unitary gate, 
a generalized form of Toffoli gate and subsequently generalize the protocol for $n$-party using minimal cost.

\end{abstract}


~~~

{\it Introduction:}
One of the most striking features of the quantum world is entanglement. This esoteric quantum property has found 
practical use in the field of quantum information \cite{1}. Qubit 
teleportation \cite{2,3}, superdense coding \cite{4}, quantum information splitting \cite{5}, secret sharing \cite{6}, remote state 
preparation \cite{7,8} and many other quantum communication protocols have been theoretically and experimentally demonstrated taking recourse to entanglement. Local implementations 
of non-local quantum gates is another quantum communication protocol implementing a multi-partite quantum gate which can 
not be decomposed into individual local operations between spatially distributed qubits. This can be achieved using 
entangled channels shared by the remote parties and local operations with classical communications (LOCC). This quantum task is 
also called gate teleportation, which is necessary for distributed quantum computing.

As is well known, controlled-NOT, together with the single qubit gate, form the universal gates to which other gates can be decomposed \cite{9}. In principle, Controlled-Unitary gates can be implemented locally, using only CNOT gate teleportation protocol. 
Involving less entanglement and communication costs, several protocols have been proposed implementing non-local multi-partite operations locally by LOCC using entangled channels \cite{10,11,12,13,14,15,16,17,18,19,20,21} and qubit communication \cite{22,23}. Probabilistic and deterministic gate implementation using non-maximally entangled state has been explicated \cite{24,25,26,27}. Assisted with linear optical manipulations, photon entanglement produced from parametric down-conversion, and post-selection from the coincidence measurements, Huang {\it et al.} \cite{28} teleported the CNOT gate experimentally. Later, other experimental protocols have been demonstrated \cite{29,30}. 
In contrast with the known protocols, we consider here an arbitrary multi-partite state, either product or entangled, where all the qubits are remote placed and demonstrate the protocol of simultaneous and $n$-qubit controlled operation in linear entangled channel.

In the familiar network by Eisert {\it et. al.} \cite{10}, each of the control parties shares one entangled state with the target party and none of the control parties shares entanglement between them whereas in linear entangled channel, the control parties share entanglement with the adjacent control parties and only one of them shares entanglement with the target party. As this network is linear, the target party has to maintain only one entangled channel, which is particularly useful when the entanglement sharing is difficult between each controlling agent with the target party. 
In this paper we start with a three party scenario, where Alice and Bob simultaneously implement controlled-Hermitian ($C^{\mathcal{H}}$) gate to Charlie in linear entangled network, which is then generalized for arbitrary multi-partite state. Next section deals with the implementation of controlled-controlled-Unitary gate and the generalization to $n$-controlled Unitary gate implementation. Finally, we conclude with directions for future work.

{\it Simultaneous Implementation of Controlled-Hermitian Gate:} 
Consider three remote parties Alice, Bob and 
Charlie possess qubits 1,4 and 7 respectively of the arbitrary state
\begin{equation}
| \psi \rangle_{147} = ( d_0 | 000 \rangle + d_1 | 001 \rangle + d_2 | 010 \rangle + d_3 | 011 \rangle + d_4 | 100 \rangle + d_5 | 101 \rangle + d_6 | 110 \rangle + d_7 | 111 \rangle )
\end{equation}
with $ \Sigma_{i=0}^{7}| d_i |^2 =1$. Now Alice and Bob want to implement Controlled-Hermitian Gate (as well as unitary) on Charlie's system simultaneously where the target qubit is common for both the control parties.
To achieve this task, Alice and Bob share a Bell state between their respective qubits 2 and 3; Bob and Charlie share a Bell state between their respective qubits 5 and 6 :
\begin{equation}
| \Phi \rangle_{23} = | \Phi \rangle_{56} = \frac{1}{\sqrt{2}}( | 00 \rangle + | 11 \rangle )
\end{equation}
Here Alice shares entanglement with another control party and Bob shares entanglement with the target party, which makes a linear entanglement connection (shown in Fig. 1).
The combined state of all the qubits possessed by Alice, Bob and Charlie is given by,
\begin{equation}
\begin{split}
| \zeta \rangle_{1234567}& = (| \psi \rangle_{147} \otimes | \Phi \rangle_{23} \otimes | \Phi \rangle_{56})\\
& = \frac{1}{2}[( d_0 | 0000000 \rangle + d_1 | 0000001 \rangle + d_2 | 0001000 \rangle + d_3 | 0001001 \rangle\\
& \quad \quad + d_4 | 1000000 \rangle+ d_5 | 1000001 \rangle + d_6 | 1001000 \rangle + d_7 | 1001001 \rangle )\\
& \quad + ( d_0 | 0000110 \rangle + d_1 | 0000111 \rangle + d_2 | 0001110 \rangle + d_3 | 0001111 \rangle\\
& \quad \quad + d_4 | 1000110 \rangle + d_5 | 1000111 \rangle + d_6 | 1001110 \rangle + d_7 | 1001111 \rangle )\\
& \quad + ( d_0 | 0110000 \rangle + d_1 | 0110001 \rangle + d_2 | 0111000 \rangle + d_3 | 0111001 \rangle\\
& \quad \quad + d_4 | 1110000 \rangle + d_5 | 1110001 \rangle + d_6 | 1111000 \rangle + d_7 | 1111001 \rangle )\\
& \quad + ( d_0 | 0110110 \rangle + d_1 | 0110111 \rangle + d_2 | 0111110 \rangle + d_3 | 0111111 \rangle\\
& \quad \quad + d_4 | 1110110 \rangle + d_5 | 1110111 \rangle + d_6 | 1111110 \rangle + d_7 | 1111111 \rangle )].
\end{split}
\end{equation}
The details protocol of simultaneous $ C^{\mathcal{H}}$ implementation through linear network is described below,

{\bf Step 1:} Alice first applies controlled-NOT gate {$C^N_{12}$} on her qubits 1 and 2.

{\bf Step 2:} Alice measures on qubit 2 in computational basis and Bob applies local operations according to the outcomes of the measurements as follows,
\begin{table}[!htbp]
\begin{center}
\begin{tabular}{|c|c|c|}
\hline
{\bf Outcomes of}&{}&{}\\
{\bf measurements}&{\bf Local operations}&{\bf Combined state after}\\
{\bf on $2$}&{\bf after measurements}&{\bf measurement and operations}\\
\hline
{}&{}&{}\\
$| 0 \rangle_{2}$&$ C^N_{35} C^N_{45}$&$ d_0 | 000000 \rangle + d_1 | 000001 \rangle + d_2 | 001100 \rangle + d_3 | 001101 \rangle$\\
$$&$$&$ + d_0 | 000110 \rangle + d_1 | 000111 \rangle + d_2 | 001010 \rangle + d_3 | 001011 \rangle$\\
$| 1 \rangle_{2}$&$ C^N_{35} C^N_{45} ~ \sigma^{3}_x$&$ + d_4 | 110100 \rangle + d_5 | 110101 \rangle + d_6 | 111000 \rangle + d_7 | 111001 \rangle$\\
$$&$$&$ + d_4 | 110010 \rangle + d_5 | 110011 \rangle + d_6 | 111110 \rangle + d_7 | 111111 \rangle$\\
\hline
\end{tabular}
\end{center}
\end{table}
 (here $\sigma_x^i, \sigma_z^i, \sigma_y^i$ are Pauli operators, with superscript `$i$' indicating the qubit operand; $C^\mathcal{U}_{mn}$ denotes controlled-Unitary gate, where $`m$' is the control bit and $`n$' is target bit)

{\bf Step 3:}  Bob measures qubit 5 in computational basis and Charlie applies local operations according to the outcomes of the measurements as follows,
\begin{table}[!htbp]
\begin{center}
\begin{tabular}{|c|c|c|}
\hline
{\bf Outcomes of}&{}&{}\\
{\bf measurements}&{\bf Local operations}&{\bf Combined state after}\\
{\bf on $5$}&{\bf after measurements}&{\bf measurement and operations}\\
\hline
{}&{}&{}\\
$| 0 \rangle_{AB}$&$C^{\mathcal{H}}_{67}$&$ d_0 | 00000 \rangle + d_1 | 00001 \rangle + d_2 | 0011 \rangle \mathcal{H} | 0 \rangle + d_3 | 0011 \rangle \mathcal{H} | 1 \rangle$\\
$$&$$&$$\\
$| 1 \rangle_{AB}$&$ C^{\mathcal{H}}_{67} ~ \sigma^{6}_x$&$ + d_4 | 1101 \rangle \mathcal{H} | 0 \rangle + d_5 | 1101 \rangle \mathcal{H} | 1 \rangle + d_6 | 11100 \rangle + d_7 | 11101 \rangle$\\
\hline
\end{tabular}
\end{center}
\end{table}

{\bf Step 4:}  Finally qubit 3 and 6 are measured in Hadamard basis by Bob and Charlie and corresponding Alice applies unitary operations to obtain the desired state,
\begin{table}[!htbp]
\begin{center}
\begin{tabular}{|c|c|c|}
\hline
{\bf Outcomes of}&{\bf}&{}\\
{\bf measurements}&{\bf Local operations}&{\bf Combined state of qubits 1,4,7}\\
{\bf on $3$ and $6$}&{\bf after measurements}&{\bf after measurement and operations}\\
\hline
$| ++ \rangle_{36}$&$\mathcal{I}$&$ $\\
$| +- \rangle_{36}$&$\sigma^{1}_z \sigma^{4}_z$&$ d_0 | 000 \rangle + d_1 | 001 \rangle + d_2 | 01 \rangle \mathcal{H} | 0 \rangle + d_3 | 01  \rangle \mathcal{H} | 1 \rangle $\\
$| -+ \rangle_{36}$&$\sigma^{1}_z$&$ + d_4 | 10 \rangle \mathcal{H} | 0 \rangle + d_5 | 10 \rangle \mathcal{H} | 1 \rangle + d_6 | 110 \rangle + d_7 | 111 \rangle$\\
$| -- \rangle_{36}$&$ \sigma^{4}_z$&$ $\\
\hline
\end{tabular}
\end{center}
\end{table}

\begin{figure*}[h]
  \begin{minipage}[b]{0.5\linewidth}
    \centering
    \includegraphics[height=7.5cm,width=9cm]{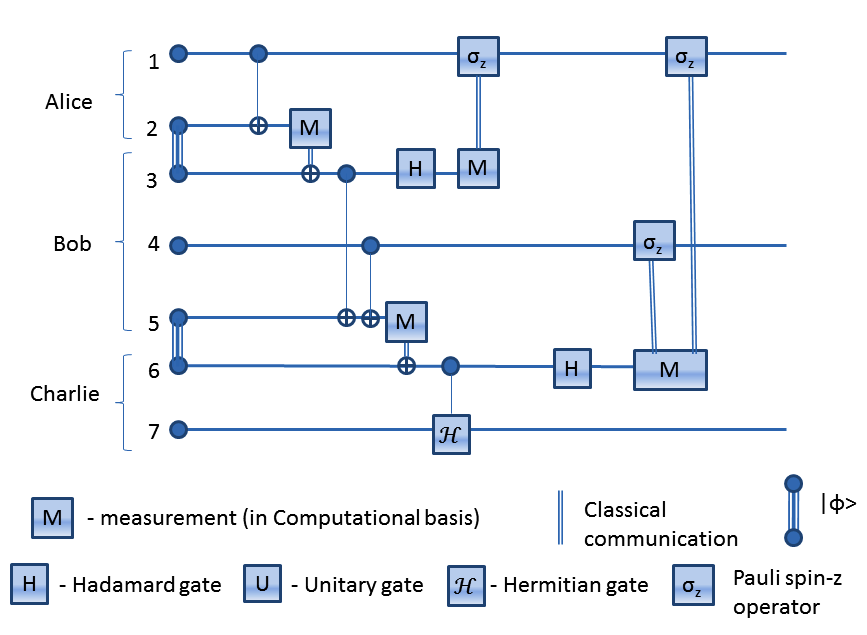}
    \caption{\it Simultaneous $C^{\mathcal{H}}$ implementation through linear network.}
  \end{minipage}
  \hspace{0.3cm}
  \begin{minipage}[b]{0.5\linewidth}
    \centering
    \includegraphics[height=8.5cm,width=10cm]{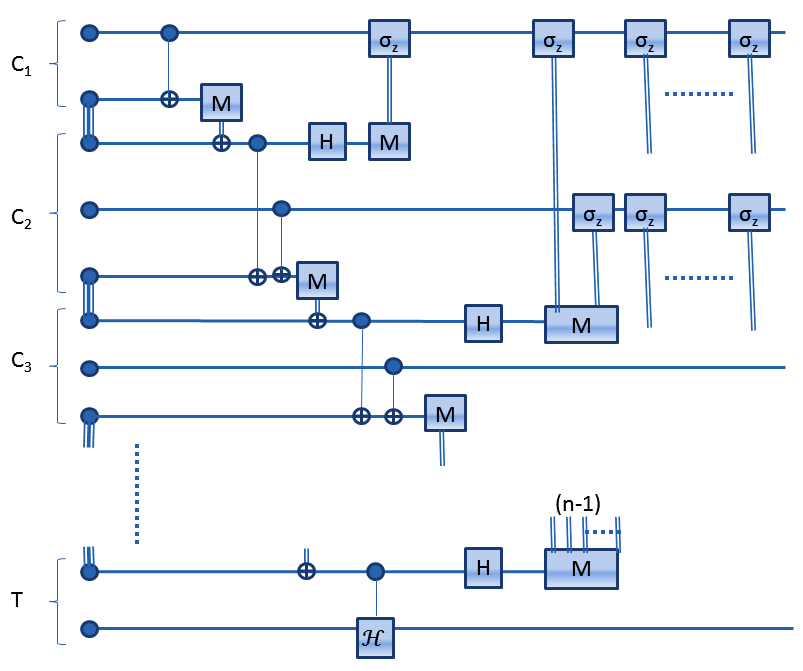}
    \caption{\it $N$ party simultaneous $C^{\mathcal{H}}$ implementation using linear network. Here $C_i$ denote the control parties and $T$ denotes the target party.}
  \end{minipage}
\end{figure*}

The pictorial representation of local unitary operations, measurements and classical communications of this protocol has been depicted in Fig. 1. The simultaneous remote implementation of controlled-Hermitian gate from two parties to one consumes 2 ebits and total 5 cbits to communicate the measurement outcomes.
The generalized protocol for $n$-party of simultaneous $ C^{\mathcal{H}} $ gate implementation described in Fig. 2, is an extension of the above protocol.
For $n$-party, the communication cost is $(n-1)$ ebits and $(n^2+n-2)/2$ cbits. 

From the above protocol, it can be inferred that the Unitary as well as Hermitian operators have significance in linear entangled network. This operator has the additional property of involution (i.e., the operator is same as its inverse), which is responsible for making this protocol deterministic. Most of the important gates like controlled-Pauli gates, controlled-Hadamard gate etc., belong to this category, making this implementation powerful.

~~~

{\it Multiparty Controlled Unitary Gate Implementation:} It has been shown that a more general form of Toffoli gate, i.e., controlled-controlled-Unitary
gate can be deterministically implemented using two Bell pairs (2 ebits of entanglement) and 4 cbits to communicate the measurement outcomes \cite{10}. Here we demonstrate the implementation of this non-local gate with the same communication cost using linear entangled channel (shown in Fig. 3). 
For illustration, we consider the same qubit distribution shared by Alice, Bob and Charlie described in Eq. 3 :
\begin{equation}
| \zeta \rangle_{1234567} = | \psi \rangle_{147} \otimes | \Phi \rangle_{23} \otimes | \Phi \rangle_{56}
\end{equation}
where we want to implement controlled-controlled-Unitary gate, $C^\mathcal{U}_{147}$ (here qubit 1 and 4 are control bits and qubit 7 is target bit) on $| \psi \rangle_{147}$. The details of the protocol is illustrated below,

{\bf Step 1}: Alice first applies controlled-NOT gate $C^N_{12}$, on her two qubits.

{\bf Step 2}: Then she measures qubit 2 in computational basis and Bob applies unitary gates as follow,
\begin{table}[h] 
\begin{center}
\begin{tabular}{|c|c|c|}
\hline
{\bf Outcomes of}&{\bf Local}&{\bf Combined state obtained}\\
{\bf measurements}&{\bf operations}&{\bf after  measurements and operations}\\
{\bf on qubit 2}&{\bf by Bob}&{\bf }\\
\hline
{}&{}&{}\\
$| 0 \rangle_2$&$ C^{N}_{345}$&$ d_0 | 000000 \rangle_ + d_1 | 000001 \rangle + d_2 | 001000 \rangle + d_3 | 001001 \rangle$\\
$$&$$&$+ d_0 | 000110 \rangle + d_1 | 000111 \rangle + d_2 | 001110 \rangle + d_3 | 001111 \rangle$\\
$$&$$&$+ d_4 | 110000 \rangle+ d_5 | 110001 \rangle + d_6 | 111100 \rangle + d_7 | 111101 \rangle$\\
$| 1 \rangle_2$&$ C^{N}_{345} \sigma^3_{x}$&$+ d_4 | 110110 \rangle + d_5 | 110111 \rangle + d_6 | 111010 \rangle + d_7 | 111011 \rangle$\\
\hline
\end{tabular}
\end{center}
\end{table}

{\bf Step 3:} Bob measures on qubit 5 in computational basis and accordingly Charlie performs unitary gates, (here $\mathcal{U} | \psi \rangle$ is denoted as $\p{| \psi \rangle}$)
\begin{table}[!htbp]
\begin{center}
\begin{tabular}{|c|c|c|}
\hline
{\bf Outcomes of}&{\bf Local}&{\bf Combined state obtained }\\
{\bf measurements}&{\bf operations}&{\bf after measurements and operations}\\
{\bf on qubit 5}&{\bf by Bob}&{}\\
\hline
{}&{}&{}\\
$| 0 \rangle_5$&$C^\mathcal{U}_{67}$&$d_0 | 00000 \rangle + d_1 | 00001 \rangle + d_2 | 00100 \rangle + d_3 | 00101 \rangle$\\
$$&$$&$+ d_4 | 11000 \rangle+ d_5 | 11001 \rangle + d_6 | 1111\p0 \rangle + d_7 | 1111\p1 \rangle$\\
$| 1 \rangle_5$&$C^\mathcal{U}_{67} \sigma_x^6$&$$\\
\hline
\end{tabular}
\end{center}
\end{table}

{\bf Step 4:} After that Charlie measures qubit 6 in Hadamard basis and Bob applies local operations depending on the outcomes,
\begin{table}[!htbp]
\begin{center}
\begin{tabular}{|c|c|c|}
\hline
{\bf Outcomes of}&{\bf Local}&{\bf Combined state obtained}\\
{\bf measurements}&{\bf operations}&{\bf after measurements and operations}\\
{\bf on qubit 6}&{\bf by Bob}&{}\\
\hline
{}&{}&{}\\
$| + \rangle_6$&$\mathcal{I}$&$d_0 | 0000 \rangle + d_1 | 0001 \rangle + d_2 | 0010 \rangle + d_3 | 0011 \rangle$\\
$$&$$&$+ d_4 | 1100 \rangle+ d_5 | 1101 \rangle + d_6 | 111\p0 \rangle + d_7 | 111\p1 \rangle$\\
$| - \rangle_6$&$C^{\sigma_{z}}_{34}$&$$\\
\hline
\end{tabular}
\end{center}
\end{table}

{\bf Step 5:} Finally qubit 3 is measured in Hadamard basis and Bob performs unitary gates to get the desired state which is shared by three parties,
\begin{table}[!htbp]
\begin{center}
\begin{tabular}{|c|c|c|}
\hline
{\bf Outcomes of}&{\bf Local}&{\bf Combined state obtained}\\
{\bf measurements}&{\bf operations}&{\bf after measurements and operations}\\
{\bf on qubit 3}&{\bf by Bob}&{}\\
\hline
{}&{}&{}\\
$| + \rangle_3$&$\mathcal{I}$&$d_0 | 000 \rangle + d_1 | 001 \rangle + d_2 | 010 \rangle + d_3 | 011 \rangle$\\
$$&$$&$+ d_4 | 100 \rangle+ d_5 | 101 \rangle + d_6 | 11\p0 \rangle + d_7 | 11\p1 \rangle$\\
$| - \rangle_3$&$\sigma_{z}^{1}$&$$\\
\hline
\end{tabular}
\end{center}
\end{table} 

\begin{figure*}[h]
  \begin{minipage}[b]{0.5\linewidth}
    \centering
    \includegraphics[height=6cm,width=9cm]{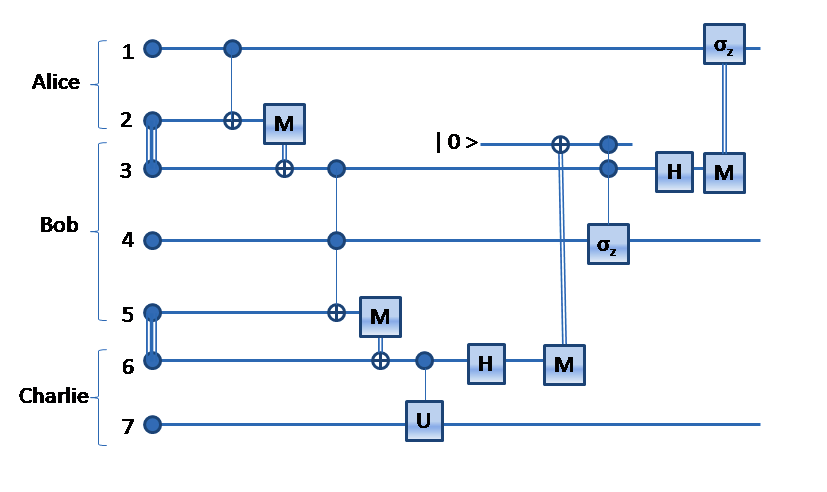}
    \caption{\it Controlled-controlled-Unitary gate implementation through linear network.}
  \end{minipage}
  \hspace{0.3cm}
  \begin{minipage}[b]{0.5\linewidth}
    \centering
    \includegraphics[height=8cm,width=10cm]{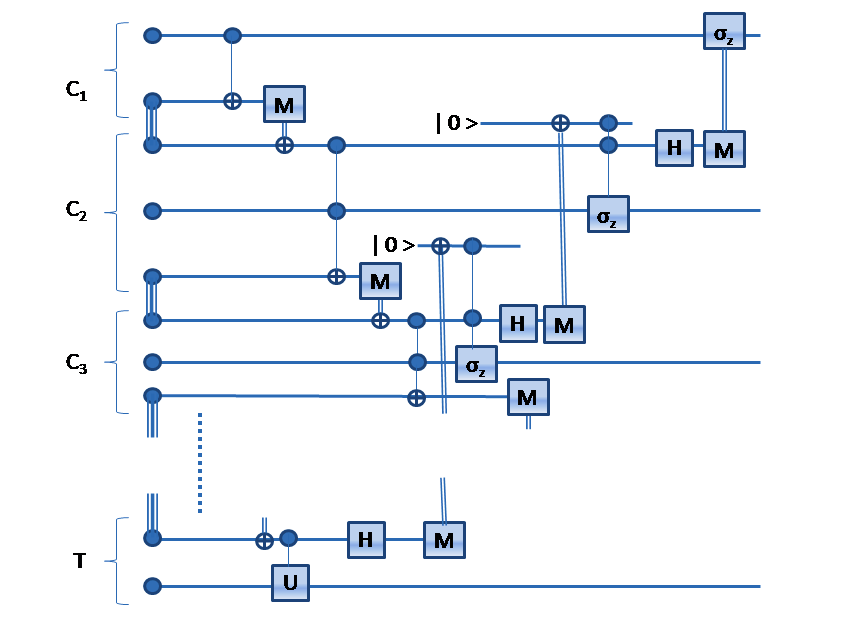}
    \caption{\it $N$ party controlled-Unitary gate implementation.}
  \end{minipage}
\end{figure*}

The above procedure can be generalized to implement a $n$-qubit
gate, where $(n-1)$ qubits are controls and the unitary operator
acts on the target qubit, only if, all the control qubits are $|1\rangle$s. The protocol is illustrated in Fig. 4 and for $n$-party gate the communication cost is $(n-1)$ ebits and $2(n-1)$ cbits which is optimal as shown in Ref. \cite{10}.

{\it Discussion:} In conclusion, we have described non-local gate implementation protocols in linear entanglement network by LOCC. 
Although the classical communication cost for implementing simultaneous controlled-gate is more as compared to the Eisert {\it et. al.} \cite{10} 
scenario, the linear network is advantageous
for large $n$, as each party shares only two entangled states and the target as well as the first control party share one entangled state. 
The fact that, our network comprises of Bell states, which are realized in laboratory conditions, makes our protocol 
experimentally achievable \cite{3}. Optimal protocol for the simultaneous 
controlled-Unitary and other non-local gate implementations in linear entangled channels can be further investigated.

{\it Acknowledgment:}
The authors acknowledge Prof. Vijay A. Singh of Homi Bhabha Centre
for Science Education (HBCSE\textemdash{}TIFR), Mumbai, India for continuous encouragement. This work is supported by
the `National Initiative on Undergraduate Science' (NIUS) program,
undertaken by HBCSE. 

We acknowledge that summary of this work was presented as a poster in `Asian Quantum Information Science Conference 2013'.

\end{document}